\documentclass[a4paper]{article}
\usepackage{mathrsfs}
\usepackage{graphicx}
\usepackage{latexsym}
\usepackage{amsmath}
\usepackage{amssymb}
\usepackage{textcomp}
\usepackage{amsbsy}
\usepackage{color}
\usepackage{graphics}

\title{\large \bf Dynamical suppression of spacetime torsion}
\author{Tanmoy Paul\footnote{E-mail address: pul.tnmy9@gmail.com}$~~~$and$~~$Soumitra SenGupta\footnote{E-mail address: tpssg@iacs.res.in}\\
Department of Theoretical Physics,\\
Indian Association for the Cultivation of Science,\\
2A $\&$ 2B Raja S.C. Mullick Road,\\
Kolkata - 700 032, India.\\[10mm]}
\date{}
\begin{document}
\maketitle

\begin{abstract}
A surprising feature of our present four dimensional universe is that its evolution appears to be governed solely by spacetime curvature without 
any noticeable effect of spacetime torsion. 
In the present paper, we give a possible explanation of this enigma through ``cosmological evolution'' of spacetime torsion 
in the backdrop of a higher dimensional braneworld scenario. Our results reveal that the torsion field may had a significant value 
at early phase of our universe, but gradually decreased with the expansion of the universe. This leads to a negligible 
footprint of torsion in our present visible universe. We also show that at an early epoch, when the amplitude of the torsion field was not suppressed, 
our universe underwent through an inflationary stage having a graceful exit within a finite time. 
To link the model with observational constraints, we also determine the spectral index for curvature perturbation ($n_s$) and 
tensor to scalar ratio ($r$) in the present context, which match with the results of $Planck$ 2018 (combining with BICEP-2 Keck-Array) data \cite{Planck}.
\end{abstract}
\newpage

\section{Introduction}
A surprising feature of the present universe is that its large scale behaviour appears to be 
controlled by one type of geometrical deformation only, namely curvature; while we notice practically no effect of another type 
of deformation, namely torsion. The most straightforward way of including torsion is to add an antisymmetric
component to the connection $\Gamma^{\alpha}_{\mu\nu}$, which is the essence of the so-called Einstein-Cartan theory \cite{hehl_report}. 
Once torsion enters into the theory in this manner, it can in principle couple with all matter fields having 
non zero spin. From dimensional argument, it can be easily shown that such interaction terms in general are of dimension 5, and are 
suppressed by the Planck mass ($M_p$), just as in the case of graviton couplings. But there has 
been no experimental evidence of the footprint of spacetime torsion on the present universe. An example is the 
Gravity Probe B experiment which was designed to estimate the precession of a gyroscope to observe any 
signature of spacetime torsion \cite{everitt}. However, all such probes, within the limit of their experimental precision, 
have consistently produced negative results and thereby disfavored the presence of the torsion 
in the spacetime geometry of our (3 + 1) dimensional visible universe \cite{lanmerzahl,mao,hehl}. Therefore the apparent torsion free universe 
indicates that the torsion field, if exists, must be severely suppressed at the present scale of the universe. 
Thus the question that naturally arises is : why are the effects of spacetime torsion are less perceptible than the spacetime curvature ? 
There is no satisfactory answer to this in the domain of four dimensional classical gravity models.\\
The proposals to remove torsion by quantum effects in 4 dimensional spacetime have appeared much before in \cite{buchbinder}, where 
the authors showed the invisibility of spacetime torsion (on the present energy scale of our universe) through the consideration of 
quantum corrected (caused by vacuum effects) gravitational action with torsion. There was also attempts to seek 
an answer to this in the context of higher dimensional braneworld models 
\cite{ssg_prl,horava,RS,kaloper,chodos,ssg_generalized,csaki}. In particular, in 
Randall-Sundrum (RS) scenario \cite{RS} which involves one extra compact spacelike dimension with $S^1/Z_2$ orbifolding along the 
extra dimension proposed a possible explanation for this suppression of spacetime torsion in four dimension. 
This kind of scenario postulates gravity in the five-dimensional ‘bulk’, 
whereas our four-dimensional universe is confined to one of the two 3-branes located at the two orbifold fixed 
points along the compact dimension. However it has been already shown that a rank-2 antisymmetric tensor field, generally known as 
Kalb-Ramond (KR) field ($B_{MN}$), can act as a source of spacetime torsion where the torsion is identified with rank-3 
antisymmetric field strength tensor $H_{MNL}$ having a relation with $B_{MN}$ as $H_{MNL}=\partial_{[M}B_{NL]}$ \cite{partha}. In the RS like 
scenario where both the gravity and the KR field propagate in the bulk, the exponential warping nature of spacetime geometry causes the 
KR field (or equivalently the torsion) to be diluted on the visible 3-brane \cite{ssg_prl,ashmita_ks,ashmita_prd,ashmita}. 
Also there is a recent work on spacetime torsion with 
antisymmetric tensor fields in higher curvature gravity model in the context of both four dimensional and five dimensional spacetime 
\cite{gomez}, where the authors showed that due to the effect of higher curvature term(s), 
the amplitude of torsion field gets suppressed in the course of the universe evolution.\\
However in the background of cosmological evolution, the suppression of spacetime torsion (on our present universe) 
sourced by Kalb-Ramond field still awaits a proper understanding. 
Furthermore one of our authors showed earlier that the amplitude of KR field may be significant and can play a relevant role 
in the early phase of the universe. This motivates to explore whether the ``dynamical evolution'' of KR field (from 
early universe) actually leads to a negligible footprint of torsion on the present universe in 
the backdrop of braneworld scenario. We also want to explore 
the ``cosmological evolution'' of KR field from very early universe to examine whether 
the universe underwent through an inflationary expansion \cite{guth,perkins, watson,linde,kinney,langlois,tp1,tp2}. 
In particular, the questions that we address in the present paper are :

\begin{itemize}
 \item How does the Kalb-Ramond field evolve from early era of our universe? Does this evolution lead 
 to an explanation of why the effect of torsion is so much weaker than that of curvature on the present visible brane?
 
 \item In such circumstance, does the four dimensional universe undergo an accelerating expansion at early epoch? If such an inflationary 
 scenario is allowed, then what is the dependence of the duration of inflation on the KR field energy density? Moreover what are the 
 values of the spectral index ($n_s$) and tensor to scalar ratio ($r$) in the present context?
\end{itemize}

The present paper serves a natural explanation of the above questions in the backdrop of Randall-Sundrum scenario. However the warped 
RS geometry in its original form is intrinsically unstable due to intervening bulk gravity. 
A popular way of stabilizing the interbrane separation (also known as modulus or radion) is via Goldberger-Wise (GW) 
mechanism \cite{GW,GW_radion} which proposes the existence of a bulk stabilizing scalar field. 
Some variants of RS model and its modulus stabilization are discussed in \cite{csaki,julien,wolfe,tp_generalized,tp_backreaction,tp_F(R),tp_bouncing,
tp_tunneling}. Following the GW mechanism, here we propose a 
dynamical stabilization method of the extra dimensional modulus field (coupled to the KR field through the effective field equations).\\
Our paper is organized as follows: the model is described in section-II, while section-III is reserved for presenting the cosmological field 
equations and their possible solutions from the perspective of four dimensional effective theory. Their implications and possible consequences 
are discussed in the remaining part of the paper.

\section{The model}
We consider a five dimensional compactified warped geometry two brane model with spacetime torsion in the bulk. In the present 
context, the source of torsion is taken as rank-2 antisymmetric Kalb-Ramond (KR) field $B_{MN}$ (where latin indices 
run from $0$ to $4$). Torsion can be identified with rank-3 antisymmetric field strength tensor $H_{MNL}$ which is related 
to the KR field as $H_{MNL}=\partial_{[M}B_{NL]}$. The spacetime is $S^1/Z_2$ orbifolded along the extra dimension, where 
the orbifolded fixed points are identified with two 3-branes. 
Considering $\phi$ as extra dimensional angular coordinate, two branes are located at $\phi=0$ (hidden brane) and at $\phi=\pi$ (visible brane) 
respectively while the latter one is identified with the visible universe. 
One of the crucial aspects of this braneworld scenario is to stabilize the distance between the branes 
(known as modulus or radion). For this purpose, one needs to generate a suitable modulus potential with a stable minima and in 
order to do this, here we consider a massive scalar field in the five dimensional bulk. Therefore the action of the model is given by, 
\begin{eqnarray}
 S&=&\int d^4xd\phi \sqrt{-G} \bigg[\frac{R^{(5)}}{2\kappa^2} - \Lambda + V_h\delta(\phi) + V_v\delta(\phi-\pi)\nonumber\\
 &-&\frac{1}{2}G^{MN}\partial_M\Psi\partial_N\Psi - \frac{1}{2}m^2\Psi^2 - \frac{1}{12}H_{MNL}H^{MNL}\bigg]
 \label{5d action}
\end{eqnarray}
where $R^{(5)}$ is the five dimensional Ricci scalar formed by the metric $G_{MN}$, $\frac{1}{2\kappa^2}=M^3$ ($M$ is the 5 dimensional 
Planck mass), $\Lambda (<0)$ is the bulk cosmological constant 
and $V_h$, $V_v$ are the brane tensions on hidden, visible brane respectively. $\Psi$ is the stabilizing scalar field with 
$m$ denoting its mass. The KR field action is represented by the last term in the above action.\\
Considering a negligible backreaction of the KR field ($B_{MN}$) and the scalar field ($\Psi$) on the background 
spacetime, the solution of metric $G_{MN}$ turns out to be same as well known RS model i.e
\begin{eqnarray}
 ds^2 = e^{-2kr_c\phi}\eta_{\mu\nu}dx^{\mu}dx^{\nu} + r_c^2d\phi^2
 \label{5d metric1}
\end{eqnarray}
where $k=\sqrt{\frac{-\Lambda}{24M^3}}$ and $r_c$ is the interbrane separation. With this metric, the scalar field equation of motion in the bulk 
is following,
\begin{eqnarray}
\frac{1}{r_c^2}\frac{\partial}{\partial\phi}\bigg[e^{-4kr_c\phi}\frac{\partial\Psi}{\partial\phi}\bigg] - m^2 e^{-4kr_c\phi}\Psi(\phi) = 0
\label{5d scalar equation}
\end{eqnarray}
where $\Psi$ is taken as the function of $\phi$ only. Considering non-zero value of $\Psi$ on the branes, the above eqn.(\ref{5d scalar equation}) 
has the general solution,
\begin{eqnarray}
 \Psi(\phi) = e^{2kr_c\phi} \bigg[Ae^{\sigma kr_c\phi} + Be^{-\sigma kr_c\phi}\bigg]
 \label{5d scalar solution1}
\end{eqnarray}
where $\sigma = \sqrt{4 + \frac{m^2}{k^2}}$. Further the integrations constants $A$ and $B$ are obtained from the boundary conditions, 
$\Psi(0)=v_h$ and $\Psi(\pi)=v_v$ as follows,
\begin{eqnarray}
 A&=&v_v e^{-(\sigma+2)kr_c\pi} - v_h e^{-2\sigma kr_c\pi}\nonumber\\
 B&=&v_h\big(1 + e^{-2\sigma kr_c\pi}\big) - v_v e^{-(\sigma+2)kr_c\pi}
 \label{A and B}
\end{eqnarray}

Using the five dimensional spacetime metric (see eqn.(\ref{5d metric1})), different components of stress 
 tensor of the stabilizing scalar field ($\Psi$) can be obtained as,
 \begin{equation}
  T_{\phi\phi}(\Psi) = \frac{1}{4}r_c^2\bigg[-\frac{1}{r_c^2}(\partial_\phi\Psi)^2 + m^2\Psi^2\bigg]
  \nonumber\\
 \end{equation}
 and 
 \begin{equation}
  T_{\mu\nu}(\Psi) = e^{-2kr_c\phi}\eta_{\mu\nu}\frac{1}{4}\bigg[\frac{1}{r_c^2}(\partial_\phi\Psi)^2 + m^2\Psi^2\bigg]
  \nonumber\\
 \end{equation}
 Putting the bulk scalar field solution (eqn. \ref{5d scalar equation}) in the expression of $T_{\phi\phi}(\Phi)$ and $T_{\mu\nu}(\Phi)$ and 
 using the form of $A$ and $B$ in terms of $v_v$ and $v_h$ (eqn. \ref{A and B}), one can show that 
 the ratio of corresponding component of stress tensor between bulk scalar field and bulk cosmological 
 constant varies as $v_v^2/M^3$ i.e 
 \begin{eqnarray}
 \bigg(\frac{T_{\phi\phi}(\Psi)}{T_{\phi\phi}(\Lambda)}\bigg) \sim v_v^2/M^3~~~~~,~~~~~~ 
 \bigg(\frac{T_{\mu\nu}(\Psi)}{T_{\mu\nu}(\Lambda)}\bigg) \sim v_v^2/M^3
 \nonumber
 \end{eqnarray}
 where $T_{\phi\phi}(\Lambda)$ and $T_{\mu\nu}(\Lambda)$ are different components of stress tensor for the bulk cosmological 
 constant. Similarly the Lagrangian density for Kalb-Ramond field leads to the ratio of KR field stress tensor with the bulk cosmological constant as 
 $\sim H_{MNL}H^{MNL}/M^5$. Thus the stress tensor 
 for the bulk scalar field as well as for the KR field is less than that of the bulk cosmological constant for $v_v^2/M^3$ and 
 $H_{MNL}H^{MNL}/M^5$ less than unity. These conditions allow us to neglect the backreaction of the stabilizing scalar field 
 and the KR field (on the background spacetime) in comparison to bulk cosmological constant.\\
In order to introduce the radion field, we consider a fluctuation of branes around the stable configuration ($r_c$). 
So, the interbrane separation can be considered as a field ($T(x)$) and here, for simplicity, we assume that this new field 
depends only on the brane coordinates. The corresponding metric ansatz is,
\begin{eqnarray}
 ds^2 = e^{-2kT(x)\phi}g_{\mu\nu}(x) dx^{\mu}dx^{\nu} + T(x)^2d\phi^2
 \label{5d metric2}
\end{eqnarray}
Correspondingly the introduction of radion field leads to the bulk scalar field ($\Psi$) solution as follows,
\begin{eqnarray}
 \Psi(x,\phi) = e^{2kT(x)\phi} \bigg[Ae^{\sigma kT(x)\phi} + Be^{-\sigma kT(x)\phi}\bigg]
 \label{5d scalar solution2}
\end{eqnarray}
where $A$ and $B$ are given by the following expressions,
\begin{eqnarray}
 A&=&v_v e^{-(\sigma+2)kT(x)\pi} - v_h e^{-2\sigma kT(x)\pi}\nonumber\\
 B&=&v_h\big(1 + e^{-2\sigma kT(x)\pi}\big) - v_v e^{-(\sigma+2)kT(x)\pi}
 \nonumber
\end{eqnarray}

Having these set-up, now we proceed to obtain the effective four dimensional action leading to a viable physical 
description of our visible universe. In the following few subsections, we individually determine the explicit form of 4D effective action 
for various parts of the original five dimensional action (eqn.(\ref{5d action})).

\subsection{Effective action for 5D Einstein-Hilbert term}
With the metric in eqn.(\ref{5d metric2}), a Kaluza-Klein reduction for the five dimensional Einstein-Hilbert action reduces 
to four dimensional effective action as,
\begin{eqnarray}
 S_{eff}^{(1)} = \int d^4x \sqrt{-g}\bigg[\frac{M^3}{k}R^{(4)} 
 - \frac{12M^3}{k}g^{\mu\nu}\partial_{\mu}\bigg(e^{-k\pi T(x)}\bigg)\partial_{\nu}\bigg(e^{-k\pi T(x)}\bigg)\bigg]
 \label{effective action non canonical}
\end{eqnarray}
where $R^{(4)}$ is the four dimensional Ricci scalar formed by the on-brane metric $g_{\mu\nu}$. As it is evident that $T(x)$ is not canonical 
and thus we redefine the field by the following transformation :
\begin{eqnarray}
 T(x) \longrightarrow \xi(x) = \sqrt{\frac{24M^3}{k}}e^{-k\pi T(x)}
 \label{transformation}
\end{eqnarray}
In terms of the canonical radion field $\xi(x)$, $S_{eff}^{(1)}$ takes the following form,
\begin{eqnarray}
 S_{eff}^{(1)} = \int d^4x \sqrt{-g}\bigg[\frac{M^3}{k}R^{(4)} 
 - \frac{1}{2}g^{\mu\nu}\partial_{\mu}\xi\partial_{\nu}\xi\bigg]
 \label{effective action1}
\end{eqnarray}

\subsection{Effective action for bulk scalar field ($\Psi$): Radion potential}
Plugging the bulk scalar field solution (see eqn.(\ref{5d scalar solution2})) back into the five dimensional scalar field action 
$S_{scalar} = \int d^4xd\phi\sqrt{-G}\big[-\frac{1}{2}G^{MN}\partial_{M}\Psi\partial_N\Psi-\frac{1}{2}m^2\Psi^2\big]$ and integrating 
over the extra dimensional coordinate $\phi$ yields the effective action as follows,
\begin{eqnarray}
 S_{eff}^{(2)} = \int d^4x\sqrt{-g}\bigg[-\frac{k^3}{144M^6}\xi^4\bigg(v_h(\xi/f)^{\sigma} - v_v\bigg)^2\bigg]
 \label{effective action2}
\end{eqnarray}

where $f=\sqrt{\frac{24M^3}{k}}$. Further to derive the above expression, we use the relation between $T(x)$ and $\xi(x)$ 
as shown in eqn.(\ref{transformation}). However it may be noticed that the integrand of eqn.(\ref{effective action2}) acts 
as a potential term for the radion field. Afterwards we denote this potential by $V(\xi)$ i.e
\begin{eqnarray}
 V(\xi) = \frac{k^3}{144M^6}\xi^4\bigg(v_h(\xi/f)^{\sigma} - v_v\bigg)^2
 \label{radion potential}
\end{eqnarray}

Eqn.(\ref{radion potential}) clearly indicates that the potential $V(\xi)$ goes to zero in absence of the bulk scalar field 
(i.e $v_h=v_v=0$). Therefore as mentioned earlier, the potential term for the radion field is generated entirely due to the presence 
of the bulk scalar field $\Psi$ \cite{GW,GW_radion}.\\
The potential in eqn.(\ref{radion potential}) has a minimum at
\begin{eqnarray}
 \xi_{min}&=&<\xi>\nonumber\\
 &=&f\bigg[\frac{v_v}{v_h}\bigg]^{1/\sigma}
 \label{minima}
\end{eqnarray}
and a maxima at
\begin{eqnarray}
 \xi_{max} = f\bigg[\frac{v_v}{v_h}\bigg(\frac{2}{2+\sigma}\bigg)\bigg]^{1/\sigma}
 \label{maxima}
\end{eqnarray}
respectively. Moreover $V(\xi)$ goes to zero as $\xi=0$. In figure [1] we give a plot of $V(\xi)$ against $\xi$. With the expression of 
$<\xi>$, we determine the squared mass of the radion field as
\begin{eqnarray}
 m_{rad}^2 = \frac{k^2v_v^2\sigma^2}{3M^3}\bigg(\frac{v_v}{v_h}\bigg)^{2/\sigma}
 \label{radion_mass}
\end{eqnarray}

\begin{figure}[!h]
\begin{center}
 \centering
 \includegraphics[width=3.5in,height=2.0in]{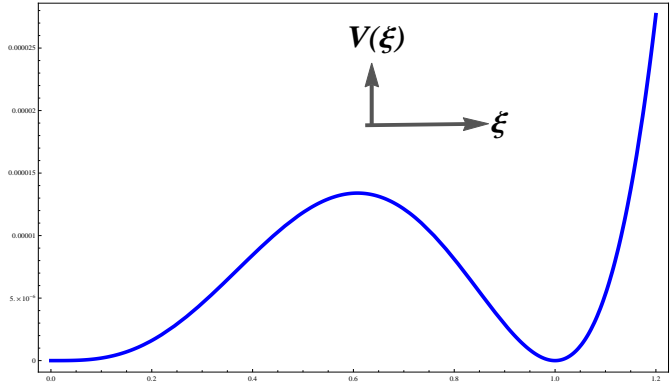}
 \caption{$V(\xi)$ vs $\xi$}
 \label{plot alpha vs brane separation}
\end{center}
\end{figure}

\subsection{Effective action for KR field action}
Recall that the 5D KR field action is given by,
\begin{eqnarray}
 S_{H} = -\frac{1}{12}\int d^4xd\phi \sqrt{-G}\bigg[H_{MNL}H^{MNL}\bigg]
 \label{KR action1}
\end{eqnarray}
where the KR field strength tensor $H_{MNL}$ is related to $B_{MN}$ (second rank antisymmetric tensor field) as 
$H_{MNL} = \partial_{[M}B_{NL]}$, 
with latin and greek indices running from $0$ to $4$ and $0$ to $3$ respectively. 
It is easy to see that the action $S[H]$ is invariant under the gauge 
transformation as $B_{MN} \rightarrow B_{MN}+\partial_{M}\omega_{N}$ (with $W_{N}$ as an arbitrary function of spacetime coordinates). This gauge 
invariance of KR field allows us to set $B_{4\mu} = 0$. Using the form of $G_{MN}$ (see eqn.(\ref{5d metric2})) and keeping $B_{4\mu}=0$, the 
above action turns out to be,
\begin{eqnarray}
 S_{H}&=&-\frac{1}{12} \int d^4xd\phi \sqrt{-g}e^{2kT(x)\phi}T(x) 
 \bigg[g^{\mu\alpha}g^{\nu\beta}g^{\lambda\gamma}H_{\mu\nu\lambda}H_{\alpha\beta\gamma}\nonumber\\
 &-&\frac{3}{T(x)^2}e^{-2kT(x)\phi}g^{\mu\alpha}g^{\nu\beta}B_{\mu\nu}\partial_{\phi}^2B_{\alpha\beta}\bigg]
 \label{KR action2}
\end{eqnarray}
The Kaluza-Klein decomposition for the KR field can be written as,
\begin{eqnarray}
 B_{\mu\nu}(x,\phi) = \sum B_{\mu\nu}^{(n)}(x)\chi^{(n)}(x,\phi)
 \label{KR decompose}
\end{eqnarray}
where $B_{\mu\nu}^{(n)}(x)$ and $\chi^{(n)}(x,\phi)$ represent the $n$th mode of on-brane KR field and extra dimensional 
KR wave function respectively. It may be mentioned that the wave function $\chi^{(n)}$ 
is considered to be a function of brane coordinates also (apart from the coordinate $\phi$), this is because our motive 
is to investigate whether the ``dynamical evolution'' of KR field leads to its invisibility on our present universe.\\
Substituting the decomposition in the 5-dimensional action $S_{H}$ and integrating over the extra dimension, 
the four dimensional effective action turns out to be:
\begin{eqnarray}
 S_{eff}^{(3)}&=&-\frac{1}{12} \int d^4x \sqrt{-g}
 \bigg[g^{\mu\alpha}g^{\nu\beta}g^{\lambda\gamma}H_{\mu\nu\lambda}^{(n)}H_{\alpha\beta\gamma}^{(n)}\nonumber\\
 &+&3m_n^2 g^{\mu\alpha}g^{\nu\beta}B_{\mu\nu}^{(n)}B_{\alpha\beta}^{(n)}\bigg]
 \label{effective action3}
\end{eqnarray}
provided $\chi^{(n)}(x,\phi)$ satisfies the following equation of motion, 
\begin{eqnarray}
 \frac{\partial\chi^{(n)}}{\partial t} \frac{\partial\chi^{(m)}}{\partial t} 
 - \frac{1}{T^2(t)}e^{-2kT(t)\phi}\chi^{(n)}\frac{\partial^2\chi^{(m)}}{\partial\phi^2} = m_n^2\chi^{(n)}\chi^{(m)}
 \label{wave function equation}
\end{eqnarray}
along with the normalization condition  as,
\begin{eqnarray}
 \int_0^{\pi} d\phi e^{2kT(t)\phi}\chi^{(n)}\chi^{(m)} = \frac{1}{T^2(t)}\delta_{mn}
 \label{wave function normalization}
\end{eqnarray}
where $m_n$ denotes the mass of nth KK mode. As we will see later that $\chi^{(n)}(x,\pi)$ is important to determine 
the coupling between the KR field and various Standard Model fields on the visible brane. 
Further eqn.(\ref{wave function equation}) clearly demonstrates that the dynamical evolution of $\chi^{(n)}(x,\phi)$ 
is coupled with the modulus (or radion) field $T(x)$.\\

Eqns.(\ref{effective action1}), (\ref{effective action2}) and (\ref{effective action3}) immediately lead to the final form of the four dimensional 
effective action as follows :
\begin{eqnarray}
 S_{eff}&=&S_{eff}^{(1)} + S_{eff}^{(2)} + S_{eff}^{(3)}\nonumber\\
 &=&\int d^4x\sqrt{-g}\bigg[\frac{M^3}{k}R^{(4)} - \frac{1}{2}g^{\mu\nu}\partial_{\mu}\xi\partial_{\nu}\xi - V(\xi)\nonumber\\
 &+&g^{\mu\alpha}g^{\nu\beta}g^{\lambda\gamma}H_{\mu\nu\lambda}^{(n)}H_{\alpha\beta\gamma}^{(n)} 
 + 3m_n^2 B_{\mu\nu}^{(n)}B^{\mu\nu(n)}\bigg]
 \label{full effective action_secondary}
\end{eqnarray}
where the radion potential $V(\xi)$ is explicitly shown in eqn.(\ref{radion potential}). At this stage it deserves mentioning that 
the zeroth Kaluza-Klein (KK) mode of the field strength tensor (i.e $H^{(0)}_{\mu\nu\lambda}$) can be identified 
with spacetime torsion and thus from now on, we deal with the zeroth mode of the KR field for which $m_{n=0} = 0$. With this 
lowest KK mode, the four dimensional effective action turns out to be,

\begin{eqnarray}
 S_{eff}&=&S_{eff}^{(1)} + S_{eff}^{(2)} + S_{eff}^{(3)}\nonumber\\
 &=&\int d^4x\sqrt{-g}\bigg[\frac{M^3}{k}R^{(4)} - \frac{1}{2}g^{\mu\nu}\partial_{\mu}\xi\partial_{\nu}\xi - V(\xi)\nonumber\\ 
 &+&g^{\mu\alpha}g^{\nu\beta}g^{\lambda\gamma}H_{\mu\nu\lambda}^{(0)}H_{\alpha\beta\gamma}^{(0)}\bigg]
 \label{full effective action}
\end{eqnarray} 

However due to the presence of the potential $V(\xi)$, 
the radion field acquires a certain dynamics governed by the effective field equations. In this  scenario, our motivation is to investigate  whether 
the dynamics of the radion field can trigger such a evolution on the KR wave function $\chi^{(0)}(x,\phi)$, that will lead 
to the fact that the effect of KR field (or equivalently the torsion field) evolves to get suppressed with the expansion of our universe. 
Further in \cite{ashmita}, it was shown that the energy density of $B^{(0)}_{\mu\nu}$ may be dominant and can have a significant role 
at early phase of the universe. Therefore it is crucial to explore the dynamical evolution of KR field from very early era of the 
universe where it is also an intriguing part to examine whether the early universe passes through an inflationary period or not. 
Motivated by this idea, we try to solve the cosmological Freidmann equations obtained from the four dimensional effective action $S_{eff}$. 
This is demonstrated in the next section.\\
At this stage it deserves mentioning that the energy scale, or the compactification scale, of the five dimensional bulk is $\sim$ Planck scale. However 
as mentioned earlier that here we are interested on inflation on our 4D visible universe, 
where the energy scale (or the inverse of the duration of inflation) comes 
with $\sim 10^{10}$ GeV which is consistent with the Planck observations as has been described later. Thus the 4D inflationary energy scale is 
lesser compared to the 5D bulk scale and we can consider the 4D effective action where the extra dimensional component of 5D metric i.e the modulus 
appears as radion field. Thus the approach here is motivated by the calculation of the effective action proposed 
by Goldberger and Wise in \cite{GW,GW_radion}.

\section{Effective cosmological equations and their possible solutions}
In order to obtain the effective field equations, 
first we determine the energy-momentum tensor for $\xi(x)$ and $B_{\mu\nu}^{(0)}(x)$ as,
\begin{eqnarray}
 T_{\mu\nu}[\xi]&=&\frac{2}{\sqrt{-g}}
 \frac{\delta}{\delta g^{\mu\nu}}\bigg[\sqrt{-g}\bigg(\frac{1}{2}g^{\alpha\beta}\partial_{\alpha}\xi\partial_{\beta}\xi + V(\xi)\bigg)\bigg]\nonumber\\
 &=&\partial_{\mu}\xi\partial_{\nu}\xi - g_{\mu\nu}\bigg(\frac{1}{2}g^{\alpha\beta}\partial_{\alpha}\xi\partial_{\beta}\xi + V(\xi)\bigg)
 \label{em tensor1}
\end{eqnarray}
and
\begin{eqnarray}
 T_{\mu\nu}[B^{(0)}]&=&\frac{2}{\sqrt{-g}}
 \frac{\delta}{\delta g^{\mu\nu}}\bigg[\frac{1}{12}\sqrt{-g}g^{\mu\alpha}g^{\nu\beta}g^{\lambda\gamma}H_{\mu\nu\lambda}H_{\alpha\beta\gamma}\bigg]\nonumber\\
 &=&\frac{1}{6}\bigg[3g_{\nu\rho}H_{\alpha\beta\mu}H^{\alpha\beta\rho} 
 - \frac{1}{2}g_{\mu\nu}H_{\alpha\beta\gamma}H^{\alpha\beta\gamma}\bigg]
 \label{em tensor2}
\end{eqnarray}
respectively.\\
The on-brane metric ansatz that fits our purpose is the flat FRW metric i.e
\begin{eqnarray}
 ds_{(4)}^2&=&g_{\mu\nu}(x) dx^{\mu}dx^{\nu}\nonumber\\
 &=&-dt^2 + a^2(t)\big[dx^2 + dy^2 + dz^2\big]
 \label{4d metric}
\end{eqnarray}
where $a(t)$ is the scale factor of the visible universe. 
However before presenting the field equations, we want to emphasize that due to antisymmetric nature, $H_{\mu\nu\lambda}^{(0)}$ has 
four independent components on the visible 3-brane, they can be expressed as,
\begin{eqnarray}
 H_{012}^{(0)} = h_1~~~~~~~~~~~~,~~~~~~~~~~~~H^{012(0)} = h^1\nonumber\\
 H_{013}^{(0)} = h_2~~~~~~~~~~~~,~~~~~~~~~~~~H^{013(0)} = h^2\nonumber\\
 H_{023}^{(0)} = h_3~~~~~~~~~~~~,~~~~~~~~~~~~H^{023(0)} = h^3\nonumber\\
 H_{123}^{(0)} = h_4~~~~~~~~~~~~,~~~~~~~~~~~~H^{123(0)} = h^4
\nonumber
\end{eqnarray}
With these independent components along with the metric shown in eqn.(\ref{4d metric}), we determine 
various components of $T_{\mu\nu}[\xi]$ and $T_{\mu\nu}[B^{(0)}]$, as given in Appendix-1. Such expressions of energy-momentum 
tensor immediately lead to the off-diagonal Friedmann equations (obtained from the effective action $S_{eff}$ in eqn.(\ref{full effective action})) as,
\begin{eqnarray}
 h_4h^3 = h_4h^2 = h_4h^1 = h_2h^3 = h_1h^3 = h_1h^2 = 0
 \label{off einstein equation}
\end{eqnarray}
The above set of equations has the following solution,
\begin{eqnarray}
 h_1 = h_2 = h_3 = 0~~~~~~~~~~~~~~~~~, h_4 \neq 0
 \label{sol off einstein equation}
\end{eqnarray}
Using this solution, one easily obtains total energy density and pressure for the matter fields ($\xi$, $B_{\mu\nu}^{(0)}$) as 
$\rho_T = \bigg[\frac{1}{2}\dot{\xi}^2 + V(\xi) + \frac{1}{2}h_4h^4\bigg]$ and 
$p_T = \bigg[\frac{1}{2}\dot{\xi}^2 - V(\xi) + \frac{1}{2}h_4h^4\bigg]$ respectively (where the fields are taken to be homogeneous in space and an 
overdot denotes $\frac{d}{dt}$). As a result, the diagonal Friedmann equations take the following form,
\begin{eqnarray}
 3H^2 = \frac{1}{2}\dot{\xi}^2 + V(\xi) + \frac{1}{2}h_4h^4
 \label{einstein equation1}
\end{eqnarray}

\begin{eqnarray}
 2\dot{H} + 3H^2 + \frac{1}{2}\dot{\xi}^2 - V(\xi) + \frac{1}{2}h_4h^4 = 0
 \label{einstein equation2}
\end{eqnarray}
where $H=\frac{\dot{a}}{a}$ is known as Hubble parameter. Further, the effective field equations for the zeroth mode of KR field ($B_{\mu\nu}^{(0)}$) 
and the radion field ($\xi$) are given by,
\begin{eqnarray}
 \nabla_{\mu}H^{\mu\nu\lambda(0)} = \frac{1}{\sqrt{-g}}\partial_{\mu}\bigg[\sqrt{-g}H^{\mu\nu\lambda(0)}\bigg] = 0
 \label{KR equation}
\end{eqnarray}
and
\begin{eqnarray}
 \ddot{\xi} + 3H\dot{\xi} + \frac{\partial V}{\partial\xi} = 0
 \label{radion equation}
\end{eqnarray}
respectively, where $V(\xi)$ is explicitly shown in eqn.(\ref{radion potential}). 
However the only information that we get from eqn.(\ref{KR equation}) is that the non-zero component of $H_{\mu\nu\lambda}^{(0)}$ 
i.e $h_4$ depends on the coordinate $t$ (see Appendix-2 for the derivation), as is also expected from the gravitational field equations. 
Taking time derivative of both sides of 
eqn.(\ref{einstein equation1}), we get $6H\dot{H}=\bigg[\dot{\xi}\ddot{\xi}+V'(\xi)\dot{\xi}+\frac{1}{2}\frac{d}{dt}(h_4h^4)\bigg]$. Further 
eqns.(\ref{einstein equation1}) and (\ref{einstein equation2}) immediately lead to an expression as 
$2\dot{H}=-\dot{\xi}^2-\frac{1}{2}h_4h^4$. Equating these two expressions of $\dot{H}$ and using the radion field equation, one finally lands with 
the following time evolution for $h_4h^4$ as,
\begin{eqnarray}
 \frac{d}{dt}(h_4h^4) = -6Hh_4h^4
 \nonumber
\end{eqnarray}
Solving the above differential equation, we obtain
\begin{eqnarray}
 h_4h^4 = \frac{h_0}{a^6(t)}
 \label{sol KR energy density}
\end{eqnarray}
where $h_0$ is an integration constant which is restricted to take only positive values in order to get a real solution of $h_4$. 
Recall that the term $\frac{1}{2}h_4h^4$ represents the energy density contributed from the KR field i.e $\rho_{KR} = \frac{1}{2}h_4h^4$. Therefore 
eqn.(\ref{sol KR energy density}) clearly indicates that the energy density of the KR field (zeroth mode) decreases monotonically as the universe 
expands with time. This leads to a negligible footprint of spacetime torsion on our present visible universe. 
However at the same time eqn.(\ref{sol KR energy density}) 
also demonstrates that the energy density of the KR field should play an important role at early phase of the universe (when $a(t)$ is small 
compared to the present one). Therefore in order to understand the dynamical suppression of the KR field, 
it is crucial to determine the time evolution of $h_4$ from very early universe where it is also important to examine 
whether the universe undergoes through an inflationary stage or not. To investigate these phenomena, 
we need to solve the scale factor during initial era.\\
Using the above form of $h_4h^4$ (see eqn.(\ref{sol KR energy density})), 
there remain two independent effective field equations,
\begin{eqnarray}
 H^2 = \frac{1}{3}\bigg[\frac{1}{2}\dot{\xi}^2 + V(\xi)\bigg] + \frac{1}{6}\frac{h_0}{a^6}
 \label{independent equation1}
\end{eqnarray}

\begin{eqnarray}
 \ddot{\xi} + 3H\dot{\xi} + \frac{\partial V}{\partial\xi} = 0
 \label{independent equation2}
\end{eqnarray}
These two equations are sufficient to determine the two unknowns namely the scale factor ($a(t)$) and the radion field ($\xi(t)$). 
As mentioned earlier, we are interested to solve eqns.(\ref{independent equation1}), (\ref{independent equation2}) during early universe 
and for this purpose, the potential energy of the radion field 
is considered to be greater than that of the kinetic energy (known as slow-roll approximation) i.e.
\begin{eqnarray}
 V(\xi) \gg \frac{1}{2}\dot{\xi}^2
 \label{slow roll approximation}
\end{eqnarray}
Under this approximation, eqn.(\ref{independent equation1}) and eqn.(\ref{independent equation2}) are simplified to,
\begin{eqnarray}
 H^2 = \frac{1}{3}V(\xi) + \frac{1}{6}\frac{h_0}{a^6}
 \label{slow roll equation1}
\end{eqnarray}
and
\begin{eqnarray}
 3H\dot{\xi} + \frac{\partial V}{\partial\xi} = 0
 \label{slow roll equation2}
\end{eqnarray}
respectively. Using the explicit form of $V(\xi)$ (see eqn. (\ref{radion potential})), we solve the above two equations 
for $\xi(t)$, $a(t)$ as,
\begin{eqnarray}
 \xi(t) = \frac{\xi_0}{\bigg[D\xi_0^{\sigma}-\big(D\xi_0^{\sigma}-\frac{\sqrt{h_0}}{a_0^3\xi_0^2}-1\big)\exp
 {\big[-\frac{1}{3}\sigma v_v\sqrt{\frac{k^3}{3M^6}}(t-t_0)\big]}\bigg]^{1/\sigma}}
 \label{sol of radion}
\end{eqnarray}
and
\begin{eqnarray}
 a(t) = C \bigg[1 + \sqrt{\frac{3h_0}{2}}(t-t_0)\bigg]^{1/3} \exp{\bigg[\frac{1}{12}v_v\sqrt{\frac{k^3}{3M^6}} \big(g_1(t)-g_2(t)\big)\bigg]} ,
 \label{sol of scale}
\end{eqnarray}
where $D = \frac{v_h}{v_v}\big(\frac{k}{24M^3}\big)^{\sigma/2}$ and $\xi_0$, $C$ are integration constants 
with $a_0=C \exp{[-\xi_0^2/8]}$. Further $g_1(t)$ has the following form,
\begin{eqnarray}
 &g_1&(t)= - \frac{D\xi_0^{\sigma}}{(D\xi_0^{\sigma}-1)} \bigg(\frac{1}{\frac{1}{3}\sigma v_v\sqrt{\frac{k^3}{3M^6}}}\bigg)\nonumber\\
 &2F1&\bigg(1,1,2+\frac{2}{\sigma},\frac{D\xi_0^{\sigma}}{D\xi_0^{\sigma}-1}
 \exp{\big(\frac{1}{3}\sigma v_v\sqrt{\frac{k^3}{3M^6}}(t-t_0)\big)}\bigg)\nonumber\\
 &\exp&{\bigg(\frac{1}{3}\sigma v_v\sqrt{\frac{k^3}{3M^6}}(t-t_0)\bigg)} \bigg(D\xi_0^{\sigma}-(D\xi_0^{\sigma}-1)\nonumber\\ 
 &\exp&{\big(-\frac{1}{3}\sigma v_v\sqrt{\frac{k^3}{3M^6}}(t-t_0)\big)}\bigg)^{-2/\sigma}
 \label{g1}
\end{eqnarray}

where $2F1$ symbolizes the hypergeometric function. Similarly the form of $g_2(t)$ is given by,

\begin{eqnarray}
 &g_2&(t)= - \frac{\xi_0^{\sigma}}{(D\xi_0^{\sigma}-1)} \bigg(\frac{1}{\frac{1}{3}\sigma v_v\sqrt{\frac{k^3}{3M^6}}}\bigg)*\nonumber\\ 
 &2F1&\bigg(1,1,1+\frac{2}{\sigma},\frac{D\xi_0^{\sigma}}{D\xi_0^{\sigma}-1}
 \exp{\big(\frac{1}{3}\sigma v_v\sqrt{\frac{k^3}{3M^6}}(t-t_0)\big)}\bigg)\nonumber\\
 &\exp&{\bigg(\frac{1}{3}\sigma v_v\sqrt{\frac{k^3}{3M^6}}(t-t_0)\bigg)} \bigg(D\xi_0^{\sigma}-(D\xi_0^{\sigma}-1)\nonumber\\
 &\exp&{\big(-\frac{1}{3}\sigma v_v\sqrt{\frac{k^3}{3M^6}}(t-t_0)\big)}\bigg)^{1-2/\sigma}.
 \label{g2}
\end{eqnarray}
It may be noticed from eqn.(\ref{sol of radion}) and eqn.(\ref{sol of scale}) that for $\Psi \rightarrow 0$ (or $v_v=0$), the solution of the radion field 
and the Hubble parameter become $\xi(t) = \frac{\xi_0}{\big[1 + \frac{\sqrt{h_0}}{a_0^3\xi_0^2}\big]^{1/\sigma}}=\xi(t_0)$ and $H \propto \frac{1}{a^3}$ 
respectively. This is expected because in the absence of bulk scalar field ($\Psi$), the potential $V(\xi)$ (see eqn.(\ref{radion potential})) 
goes to zero and thus the radion field has no dynamics which in turn makes the variation of the Hubble parameter as $H \propto \frac{1}{a^3}$ 
(solely due to the KR field having equation of state parameter $=1$).\\
Further eqn.(\ref{sol of radion}) clearly indicates that 
$\xi(t)$ decreases with time. Comparison of eqn.(\ref{minima}) and eqn.(\ref{sol of radion}) reveals that the radion 
field reaches at its vacuum expectation value (vev) asymptotically (within the slow roll approximation) at large time ($t\gg t_0$) i.e. 
\begin{eqnarray}
 \Psi(t\gg t_0)&=&f\bigg[\frac{v_v}{v_h}\bigg]^{1/\sigma}\nonumber\\
 &=&<\Psi>
\end{eqnarray}
This vev of radion field leads to the stabilized interbrane separation (between Planck and TeV branes) as,
\begin{eqnarray}
k\pi <T(x)> = \frac{4k^2}{m^2}[\ln{(\frac{v_h}{v_v})}]
 \label{brane separation}
\end{eqnarray}
where $m^2$ is squared mass of the stabilizing scalar field $\Psi$. \\

\section{Beginning of inflation}
After obtaining the solution of $a(t)$ (in eqn.(\ref{sol of scale})),   
we can now examine whether this form of scale factor corresponds to an accelerating era of the early universe (i.e. $t\gtrsim t_0$)
or not. In order to check this, we expand $a(t)$ in the form of Taylor series (about $t=t_0$) and 
retain the terms only up to first order in $t-t_0$:
\begin{eqnarray}
 a(t\gtrsim t_0)=a_0 \bigg[1 + \sqrt{\frac{3h_0}{2}}(t-t_0)\bigg]^{1/3} 
 \exp{\bigg[\frac{1}{12}\xi_0^2v_v\sqrt{\frac{k^3}{3M^6}}(D\xi_0^{\sigma}-1)(t-t_0)\bigg]}
 \label{limiting scale factor}
\end{eqnarray}
where $a_0$ is the value of the scale factor at $t=t_0$ and related to the integration constant $C$ as,
\begin{equation}
 a_0 = C \exp{[-\xi_0^2/8]}.
 \nonumber
\end{equation}
Eqn.(\ref{limiting scale factor}) leads to the acceleration of the universe at $t \rightarrow t_0$ as follows:
\begin{eqnarray}
 \frac{\ddot{a}}{a}(t\gtrsim t_0&)&= \bigg[\frac{\xi_0^2v_v}{12}\sqrt{\frac{k^3}{3M^6}}(D\xi_0^{\sigma}-1) 
 + \sqrt{\frac{h_0}{2}}\big(1+\frac{1}{\sqrt{3}}\big)\bigg]\nonumber\\ 
 &\bigg[&\frac{\xi_0^2v_v}{12}\sqrt{\frac{k^3}{3M^6}}(D\xi_0^{\sigma}-1) 
 - \sqrt{\frac{h_0}{2}}\big(1-\frac{1}{\sqrt{3}}\big)\bigg]
 \label{limiting acceleration}
\end{eqnarray}
It may be noticed that for the condition
\begin{eqnarray}
\frac{\xi_0^2v_v}{12}\sqrt{\frac{k^3}{3M^6}}(D\xi_0^{\sigma}-1) > \sqrt{\frac{h_0}{2}}\big(1-\frac{1}{\sqrt{3}}\big)
\label{condition}
\end{eqnarray}
the early universe undergoes through an accelerating stage while for 
$\frac{\xi_0^2v_v}{12}\sqrt{\frac{k^3}{3M^6}}(D\xi_0^{\sigma}-1) < \sqrt{\frac{h_0}{2}}\big(1-\frac{1}{\sqrt{3}}\big)$, $\ddot{a}(t\rightarrow t_0)$ 
becomes less than zero.\\
At this stage, it deserves mentioning that the parameters $v_v$ and $h_0$ controls the strength of the radion field and the KR field 
energy density respectively. Therefore the interplay between the radion field and the KR field fixes whether the early universe evolves through an 
accelerating stage or not. However in order to solve the flatness and horizon problems (for a review, we refer to \cite{perkins,watson}), 
the universe must passes through an accelerating stage 
at early epoch and from this requirement, here we stick to the condition shown in eqn.(\ref{condition}).\\

\section{End of inflation and reheating}
In the previous section, we show that the very early universe expands with an acceleration and this accelerating stage is  
termed as the inflationary epoch. In this section, we check whether such acceleration of the scale factor 
has an end in a finite time or not.\\
The end point of an inflationary era is defined by,
\begin{eqnarray}
 \frac{\ddot{a}}{a} = \dot{H} + H^2 = 0        
 \label{end of inflation}
\end{eqnarray}
We now examine whether this condition is consistent with the field equations shown in eqn.(\ref{slow roll equation1}) 
and eqn.(\ref{slow roll equation2}). Near the end of inflation, one can safely neglect the term 
proportional to $1/a^6$ and thus eqn.(\ref{slow roll equation1}) takes the following form (at end regime of inflation):
\begin{eqnarray}
 H^2&=&\frac{1}{3}V(\xi)\nonumber\\
 &=&\frac{k^3}{432M^6}v_v^2\xi^4 \bigg(D\xi^{\sigma} - 1\bigg)^2
 \nonumber
\end{eqnarray}
Differentiating both sides of this equation with respect to t, we get the time derivative 
of the Hubble parameter as follows,
\begin{eqnarray}
 \dot{H} = -\frac{k^3}{54M^6}v_v^2\xi^2\bigg(D\xi^{\sigma} - 1\bigg)^2
 \label{time derivative of hubble}
\end{eqnarray}
where we use the equation of the radion field ($3H\dot{\xi}+V'(\xi)=0$). Plugging back the expressions 
of $H^2$ and $\dot{H}$ into eqn.(\ref{end of inflation}), one gets the following condition on radion field, 
\begin{equation}
 \xi = 2\sqrt{2} = \xi_f = \xi(t_f)
 \label{end value of radion field}
\end{equation}
where $t_f$ is the time when the radion field acquires the value $2\sqrt{2}$ (in Planckian unit). 
Eqn. (\ref{end value of radion field}) clearly indicates that the inflationary era of the universe continues as long as the radion field 
remains greater than $\xi_f$ ($= 2\sqrt{2}$). Correspondingly the duration of inflation (i.e. $t_f-t_0$) can 
be calculated from the solution of $\xi_(t)$ as follows,
\begin{eqnarray}
 \xi_f^{\sigma} = \frac{\xi_0^{\sigma}}{\bigg[D\xi_0^{\sigma}-(D\xi_0^{\sigma}-\frac{\sqrt{h_0}}{a_0^3\xi_0^2}-1)\exp
{[-\frac{1}{3}\sigma v_v\sqrt{\frac{k^3}{3M^6}}(t_f-t_0)]}\bigg]}
 \nonumber
\end{eqnarray}
Simplifying the above expression, we obtain
\begin{eqnarray}
 t_f-t_0 = \bigg(\frac{3}{\sigma v_v\sqrt{\frac{k^3}{3M^6}}}\bigg)
 \ln\bigg[\frac{D\xi_0^{\sigma} - 1 - \frac{\sqrt{h_0}}{a_0^3\xi_0^2}}
 {D\xi_0^{\sigma} - \frac{\xi_0^{\sigma}}{\xi_f^{\sigma}}}\bigg]
 \label{duration}
\end{eqnarray}
recall $D = \frac{v_h}{v_v}\big(\frac{k}{24M^3}\big)^{\sigma/2}$ and $\sigma = \frac{m^2}{4k^2}$.\\
Therefore it is clear that the inflation comes to an end in a finite time. In order to estimate the duration of inflation explicitly, one needs the 
value of the parameters $h_0$, $\xi_0$ and $v_v$, which can be determined from the expressions of spectral index and tensor to scalar ratio as 
discussed in the next section.\\
However before moving to the next section, here we discuss the reheating in the present context and the possible effects 
of KR field (or equivalently the spacetime torsion) on it. 
Needless to say that reheating describes the production of Standard Model matter at the end of the period of accelerated expansion. For 
this purpose, we consider an example where the radion field (i.e the inflaton) is coupled to another scalar field $\zeta$, 
given by the interaction Lagrangian,
\begin{eqnarray}
 L_{int} = -g\lambda \xi\zeta^2
 \label{reheating1}
\end{eqnarray}

where $g$ is a dimensionless coupling constant and $\lambda$ is a mass scale. With this interaction Lagrangian, the decay rate of the inflaton into 
$\zeta$ particles becomes

\begin{eqnarray}
 \Gamma = \frac{g^2\lambda^2}{8\pi m_{rad}}
 \label{reheating2}
\end{eqnarray}

recall that $m_{rad}$ is the mass of the radion field (see eqn.(\ref{radion_mass})). 
Generally the energy loss of the inflaton due to the production of $\zeta$ particles is taken into 
account by adding a damping term to the inflaton equation of motion as,
\begin{eqnarray}
 \ddot{\xi} + 3H\dot{\xi} + \Gamma \dot{\xi} + \frac{\partial V}{\partial\xi} = 0
 \label{reheating3}
\end{eqnarray}

Eqn.(\ref{reheating3}) clearly indicates that the radion field losses energy due to the expansion of the universe and due to transfer to 
the $\zeta$ particles, accounted by the damping terms $3H\dot{\xi}$ and $\Gamma \dot{\xi}$ respectively. As a result the production of $\zeta$ particles 
becomes effective when the Hubble parameter becomes less or comparable to $\Gamma$, otherwise the energy loss into particles is negligible 
compared to the energy loss due to the expansion of space as occurred during early phase of the inflation. Therefore the time scale $t_h$ (let 
us call it the reheating time) after when the production of $\zeta$ becomes effective is given by
\begin{eqnarray}
 H(t_h) = \Gamma
 \label{reheating4}
\end{eqnarray}

With the solution of scale factor (see eqn.(\ref{sol of scale})), the above equation turns out to be,

\begin{eqnarray}
\frac{1}{12}v_v\sqrt{\frac{k^3}{3M^6}} \big(\dot{g}_1(t_h) - \dot{g}_2(t_h)\big)  
+ \frac{\sqrt{\frac{3h_0}{2}}}{3\bigg(1 + \sqrt{\frac{3h_0}{2}}t_h\bigg)} = \Gamma
\label{reheating5}
\end{eqnarray}

where $g_1$ and $g_2$ are shown in eqns.(\ref{g1}) and (\ref{g2}) respectively. Recall, $h_0$ represents the energy density of the KR field 
during early universe and the presence of $h_0$ in the above expression entails that the KR field indeed affects the reheating time $t_h$. 
In order to understand the effect of KR field more clearly, we write $t_h = t_h^{(0)} + \delta t$, where $t_h^{(0)}$ is the reheating time in absence 
of KR field ($h_0 = 0$) i.e
\begin{eqnarray}
 \frac{1}{12}v_v\sqrt{\frac{k^3}{3M^6}} \big(\dot{g}_1(t_h^{(0)}) - \dot{g}_2(t_h^{(0)})\big) = \Gamma
\label{reheating6}
\end{eqnarray}

Thus $\delta t$ is the deviation of reheating time from $t_h^{(0)}$ solely due to the presence of the KR field. 
Expanding eqn.(\ref{reheating5}) in terms of $t_h = t_h^{(0)} + \delta t$, we get the following expression of $\delta t$

\begin{eqnarray}
 \delta t = -\frac{\sqrt{\frac{3h_0}{2}}}
 {\frac{1}{4}v_v\sqrt{\frac{k^3}{3M^6}}\bigg(1 + \sqrt{\frac{3h_0}{2}}t_h^{(0)}\bigg)\bigg(\ddot{g}_1(t_h^{(0)}) - \ddot{g}_2(t_h^{(0)})\bigg)}
 \label{reheating7}
\end{eqnarray}

where we use eqn.(\ref{reheating6}) and retain up to the term first order in $\delta t$. Clearly $\delta t$ becomes zero as $h_0 \rightarrow 0$, 
as expected. Using the explicit expressions of $g_1$, $g_2$ along with the 
condition $\frac{m}{k} < 1$ (i.e ratio of bulk scalar field mass to bulk curvature is less than unity, which is also consistent with Planck observations 
as described in the next section), we determine the term $\ddot{g}_1 - \ddot{g}_2$ (sitting in the denominator of eqn.(\ref{reheating7})) as follows:

\begin{eqnarray}
 \ddot{g}_1 - \ddot{g}_2&=&\bigg(\frac{D\xi_0^{\sigma}}{(D\xi_0^{\sigma}-1)}\bigg)^2 \bigg(\frac{\frac{1}{3}\sigma v_v\sqrt{\frac{k^3}{3M^6}}}
 {\big(1+2/\sigma\big)\big(2+2/\sigma\big)}\bigg) 
 \bigg(D\xi_0^{\sigma}-(D\xi_0^{\sigma}-1)\exp{\big(-\frac{1}{3}\sigma v_v\sqrt{\frac{k^3}{3M^6}}(t-t_0)\big)}\bigg)^{-2/\sigma}\nonumber\\
 &\bigg[&3~2F1\bigg(2,2,\frac{2}{\sigma},\frac{D\xi_0^{\sigma}}{D\xi_0^{\sigma}-1}
 \exp{\big(\frac{1}{3}\sigma v_v\sqrt{\frac{k^3}{3M^6}}(t-t_0)\big)}\bigg) 
 \exp{\bigg(\frac{2}{3}\sigma v_v\sqrt{\frac{k^3}{3M^6}}(t-t_0)\bigg)}\nonumber\\ 
 &+&\frac{8}{\big(3+2/\sigma\big)} 2F1\bigg(3,3,\frac{2}{\sigma},\frac{D\xi_0^{\sigma}}{D\xi_0^{\sigma}-1}
 \exp{\big(\frac{1}{3}\sigma v_v\sqrt{\frac{k^3}{3M^6}}(t-t_0)\big)}\bigg) 
 \exp{\bigg(\sigma v_v\sqrt{\frac{k^3}{3M^6}}(t-t_0)\bigg)}\bigg]
 \label{reheating8}
\end{eqnarray}

Thus the term $\ddot{g}_1 - \ddot{g}_2$ is positive. As a result, eqn.(\ref{reheating7}) immediately leads to the condition $\delta t < 0$ which in turn 
makes $t_h < t_h^{(0)}$. Thereby the presence of Kalb-Ramond field makes the reheating time lesser in comparison to the case when the KR field is absent. 
However, this is expected because the KR field corresponds to a deceleration of the universe i.e 
due to the appearance of KR field the Hubble parameter ($H(t)$) decreases with a faster rate by which $H(t)$ reaches to $\Gamma$ more quickly 
relative to the situation where the KR field is absent.

\section{Spectral index, tensor to scalar ratio and number of e-foldings}
In order to test the broad inflationary paradigm as well as particular models 
against precision observations \cite{Planck}, we need to calculate the value of spectral index($n_s$) and tensor to scalar ratio ($r$) and 
for this purpose, here we define a dimensionless parameter (known as slow roll parameter) as,
\begin{eqnarray}
 \epsilon = -\dot{H}/H^2
 \label{slow roll parameter1}
\end{eqnarray}
Recall the slow roll equation, $H^2 = \frac{1}{3}V(\xi) + \frac{h_0}{6a^6}$. Differentiating both sides of this equation with respect to time, we 
get
\begin{eqnarray}
 2\dot{H} = -\frac{1}{9H^2}\bigg(\frac{\partial V}{\partial\xi}\bigg)^2 - \frac{h_0}{a^6}
 \nonumber
\end{eqnarray}
where we use the field equation for radion field. These expressions of $\dot{H}$ and $H^2$ lead to the slow roll parameter $\epsilon$ as follows,
\begin{eqnarray}
 \epsilon = \frac{1}{2}\bigg[\frac{16p^2v_v^4\xi^6(D\xi^{\sigma}-1)^4 + \frac{3h_0}{a^6}\bigg(pv_v^2\xi^4(D\xi^{\sigma}-1)^2+\frac{h_0}{2a^6}\bigg)}
 {\bigg(pv_v^2\xi^4(D\xi^{\sigma}-1)^2 + \frac{h_0}{2a^6}\bigg)^2}\bigg]\nonumber\\
 \label{slow roll parameter2}
\end{eqnarray}
where $p = \frac{k^3}{144M^6}$ and $D = \frac{v_h}{v_v}\big(\frac{k}{24M^3}\big)^{\sigma/2}$.\\
The spectral index  and tensor to scalar ratio are defined by,
\begin{eqnarray}
 n_s&=&1 - 2\epsilon\bigg|_{t=t_0} - \frac{\dot{\epsilon}}{H\epsilon}\bigg|_{t=t_0}\nonumber\\
 r&=&16\epsilon\bigg|_{t=t_0}
 \nonumber
\end{eqnarray}
With the expression of $\epsilon$ obtained in eqn.(\ref{slow roll parameter2}), $r$ and $n_s$ turn out to be,
\begin{eqnarray}
 r = 8\bigg[\frac{16p^2v_v^4\xi_0^6(D\xi_0^{\sigma}-1)^4 + \frac{3h_0}{a_0^6}\bigg(pv_v^2\xi_0^4(D\xi_0^{\sigma}-1)^2+\frac{h_0}{2a_0^6}\bigg)}
 {\bigg(pv_v^2\xi_0^4(D\xi_0^{\sigma}-1)^2 + \frac{h_0}{2a_0^6}\bigg)^2}\bigg]\nonumber\\
 \label{ratio}
\end{eqnarray}
and
\begin{eqnarray}
 n_s = 1 - \frac{U_1}{U_2}
 \label{spectra index}
\end{eqnarray}
where $U_1$ and $U_2$ have the following expressions:
\begin{eqnarray}
 U_1&=&\bigg[384p^3v_v^6\xi_0^8(D\xi_0^{\sigma}-1)^6 
 + \frac{18h_0}{a_0^6}\bigg(pv_v^2\xi_0^4(D\xi_0^{\sigma}-1)^2 + \frac{h_0}{2a_0^6}\bigg)^2\nonumber\\ 
 &-&\frac{6h_0}{a_0^6}\bigg(16p^2v_v^4\xi_0^6(D\xi_0^{\sigma}-1)^4 + \frac{3h_0}{a_0^6}
 \bigg(pv_v^2\xi_0^4(D\xi_0^{\sigma}-1)^2 + \frac{h_0}{2a_0^6}\bigg)\bigg)\nonumber\\
 &-&\frac{144h_0}{a_0^6}p^2v_v^4\xi_0^6(D\xi_0^{\sigma}-1)^4\bigg]
 \nonumber
\end{eqnarray}
and
\begin{eqnarray}
 U_2&=&\bigg(pv_v^2\xi_0^4(D\xi_0^{\sigma}-1)^2 + \frac{h_0}{2a_0^6}\bigg)\bigg(16p^2v_v^4\xi_0^6(D\xi_0^{\sigma}-1)^4\nonumber\\
 &+&\frac{3h_0}{a_0^6}\bigg(pv_v^2\xi_0^4(D\xi_0^{\sigma}-1)^2 + \frac{h_0}{2a_0^6}\bigg)\bigg)
 \nonumber
\end{eqnarray}
respectively. It may be observed that the spectral index and tensor to scalar ratio depend on the parameters $v_v$, $h_0$ and $\xi_0$. 
To fix these parameters, 
we use the observational results of Planck 2018 ( combining with 
BICEP-2 Keck-Array data ) \cite{Planck} which put a constraint on $n_s$ and $r$ as 
$n_s = 0.9649 \pm 0.0042$ and $r < 0.064$ respectively. Here we take,
\begin{eqnarray}
\kappa v_v = \frac{\sqrt{h_0}}{M^2} \simeq 10^{-7}.
\nonumber
\end{eqnarray}
It may be mentioned that these values of $v_v$ and $h_0$ are consistent with the condition that is necessary for neglecting the 
backreaction of the bulk scalar field and the KR field on the background five dimensional spacetime.\\
Using eqns.(\ref{spectra index}), (\ref{ratio}) along with the values of $v_v$ and $h_0$, 
we give the plots (see Figure[2], Figure[3]) of $n_s$, $r$ with respect to $\xi_0$.\\

\begin{figure}[!h]
\begin{center}
 \centering
 \includegraphics[width=3.5in,height=2.5in]{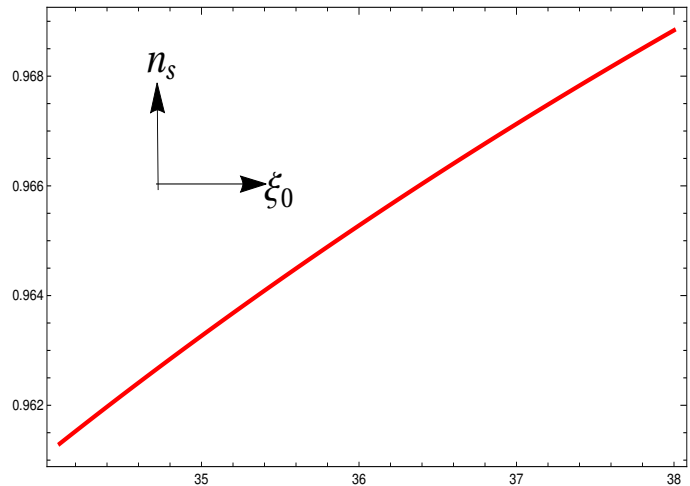}
 \caption{$n_s$ vs $\xi_0$}
 \label{plot spectral index}
\end{center}
\end{figure}

\begin{figure}[!h]
\begin{center}
 \centering
 \includegraphics[width=3.5in,height=2.5in]{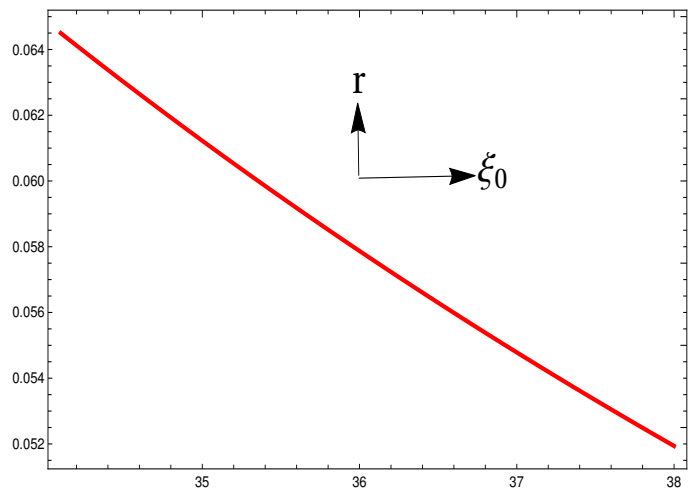}
 \caption{$r$ vs $\xi_0$}
 \label{plot tensor to scalar ratio}
\end{center}
\end{figure}

Figures [\ref{plot spectral index}] and [\ref{plot tensor to scalar ratio}] 
clearly demonstrate that for $34<\xi_0<38$ (in Planckian unit), both the observable quantities $n_s$ and $r$ remain 
within the constraints provided by $Planck$ 2018 \cite{Planck}.\\
Further with the estimated values of $v_v$, $h_0$ and $\xi_0$, the duration of inflation ($t_f-t_0$, see eqn.(\ref{duration})) comes 
as $10^{-10}$(Gev)$^{-1}$ if the ratio $m/k$ (bulk scalar field mass to bulk curvature ratio) is taken as $0.2$ \cite{GW}. We also determine the number of 
e-foldings, defined by $N = \int_{0}^{\vartriangle t}H dt$ ($\vartriangle t = t_f-t_0$, duration of inflation), 
numerically and lands with $N \simeq 58$ (with $\xi_0 = 36$, in Planckian unit).\\
In table[1], we now summarize our results:\\

\begin{table}[!h]
 \centering
\resizebox{\columnwidth}{2.0 cm}{%
  \begin{tabular}{|c| c|}
   \hline \hline
   Parameters & Estimated values\\
   \hline
   $n_s$ & $0.969$\\ 
   $r$ & 0.100\\
   $t_f-t_0$ & $10^{-10}$(GeV)$^{-1}$\\
   $N$ & 58\\
   \hline
  \end{tabular}%
  }
  \caption{Estimated values of various quantities for $\kappa v_v = \frac{\sqrt{h_0}}{M^2} \simeq 10^{-7}$ and $\xi_0 = 36$}
  \label{Table-1}
 \end{table}
 Table[\ref{Table-1}] clearly indicates that the present model may well explain the inflationary scenario of the universe 
 in terms of the observable quantities $n_s$ and $r$ as per the results of $Planck$ 2018.\\
 Using the solutions of $\xi(t)$, $a(t)$ (see eqns.(\ref{sol of radion}), (\ref{sol of scale})) along with the estimated values 
 of the parameters ($v_v$, $h_0$, $\xi_0$), we give the plots for the interbrane separation ($T(t)$, see figure[4]) and the deceleration 
 parameter ($q=-\ddot{a}/a$, see figure[5]) against a dimensionless time variable $\tilde{t} = \frac{t}{t_f}N$.\\
 
 \begin{figure}[!h]
\begin{center}
 \centering
 \includegraphics[width=3.5in,height=2.0in]{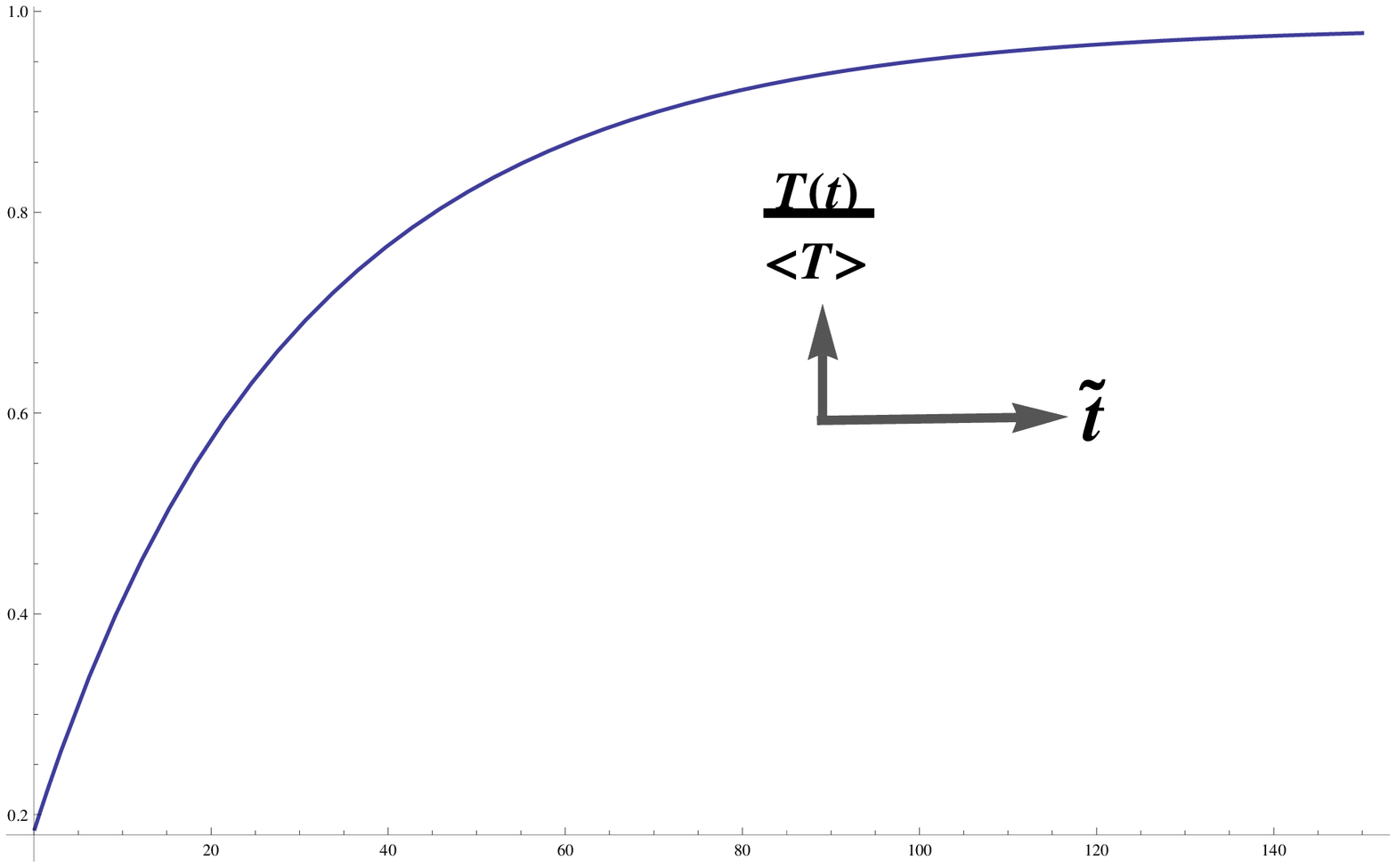}
 \caption{$\frac{T(t)}{<T>}$ vs $\tilde{t}$}
 \label{plot brane separation}
\end{center}
\end{figure}

\begin{figure}[!h]
\begin{center}
 \centering
 \includegraphics[width=3.5in,height=2.0in]{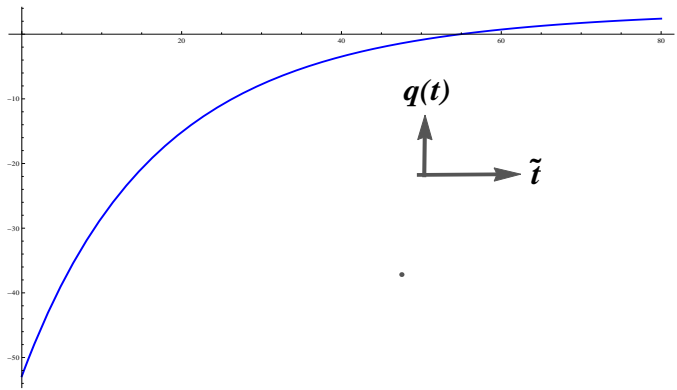}
 \caption{$q(t)$ vs $\tilde{t}$}
 \label{plot deceleration parameter}
\end{center}
\end{figure}
 
 where we use the relation $\xi(t)=\sqrt{\frac{24M^3}{k}}e^{-k\pi T(t)}$. Fig.[\ref{plot brane separation}] 
 clearly reveals that the interbrane separation 
 increases with time and saturates at $\frac{4k^2}{m^2}\ln{[v_h/v_v]}$ ($= k\pi<T>$, see eqn.(\ref{brane separation})) 
 asymptotically. It may be mentioned that 
 for $v_h/v_v = 1.5$ and $m/k = 0.2$, $k\pi<T>$ acquires the value $\simeq 36$ - required for solving the gauge hierarchy problem. Further 
 Fig.[\ref{plot deceleration parameter}] demonstrates that the early universe starts from an accelerating stage with a graceful 
 exit at a finite time.\\
 
 \section{Solution for Kalb-Ramond extra dimensional wave function}
The equation for the zeroth mode of KR wave function ($\chi^{(0)}(t,\phi)$) follows from eqn.(\ref{wave function equation}) and given by,
 \begin{eqnarray}
 \bigg(\frac{\partial\chi^{(0)}}{\partial t}\bigg)^2 
 - \frac{1}{T^2(t)}e^{-2kT(t)\phi}\chi^{(0)}\frac{\partial^2\chi^{(0)}}{\partial\phi^2} = 0
 \label{zeroth mode wave function equation}
\end{eqnarray}
As we may notice that the dynamics of the interbrane separation controls the evolution of $\chi^{(0)}(t,\phi)$.\\
It may be mentioned that the overlap of $\chi^{(0)}(t,\phi)$ with the brane $\phi=\pi$ (i.e. $\chi^{(0)}(t,\pi)$) regulates the 
coupling strengths between KR field and various Standard Model fields on the visible brane. These interaction terms play the key role 
to determine the observable signatures of KR field on our universe and thus we are interested to solve eqn.(\ref{zeroth mode wave function equation}) 
in the vicinity of $\phi=\pi$ (i.e. near the visible brane). Near the regime of $\phi \simeq \pi$, 
eqn.(\ref{zeroth mode wave function equation}) can be written as,
 \begin{eqnarray}
 \bigg(\frac{\partial\chi_v^{(0)}}{\partial t}\bigg)^2 
 - \frac{1}{T^2(t)}e^{-2k\pi T(t)}\chi_v^{(0)}\frac{\partial^2\chi_v^{(0)}}{\partial\phi^2} = 0
 \label{zeroth mode wave function equation1}
\end{eqnarray}
where $\chi_v^{(0)}$ denotes the KR wave function near the visible brane. Eqn.(\ref{zeroth mode wave function equation1}) can be solved by the 
method of separation of variables as $\chi_v^{(0)}(t,\phi) = f_1(t)f_2({\phi})$. With this expression, eqn.(\ref{zeroth mode wave function equation1}) 
turns out to be,
\begin{eqnarray}
 T^2(t)e^{2k\pi T(t)} \frac{1}{f_1^2}\bigg(\frac{df_1}{dt}\bigg)^2 = \frac{1}{f_2}\frac{d^2f_2}{d\phi^2}
 \label{zeroth mode wave function equation2}
\end{eqnarray}
As it is evident that the left and right hand side of eqn.(\ref{zeroth mode wave function equation2}) are functions of time and $\phi$ alone respectively. 
Therefore both sides of eqn.(\ref{zeroth mode wave function equation2}) can be separately equated with a constant as follows:
\begin{eqnarray}
 T^2(t)e^{2k\pi T(t)} \frac{1}{f_1^2}\bigg(\frac{df_1}{dt}\bigg)^2 = b^2
 \label{separation equation1}
\end{eqnarray}
and
\begin{eqnarray}
 \frac{1}{f_2}\frac{d^2f_2}{d\phi^2} = b^2
 \label{separation equation2}
\end{eqnarray}
where $b$ is the constant of separation. Solution of eqn.(\ref{separation equation2}) is given by $f_2(\phi) = e^{-b\phi}$, while 
eqn.(\ref{separation equation1}) is solved numerically as shown in Fig.[5] (plotted with respect to the dimensionless time variable 
$\tilde{t}=\frac{t}{t_f}N$, with $N$ be the number of e-foldings of the inflationary era).\\

\begin{figure}[!h]
\begin{center}
 \centering
 \includegraphics[width=3.5in,height=2.0in]{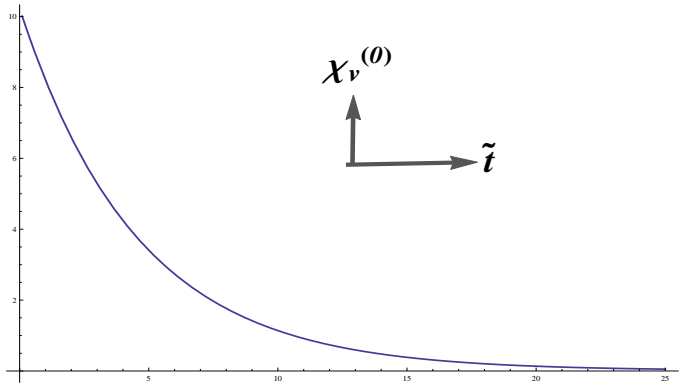}
 \caption{$\chi_v^{(0)}$ vs $\tilde{t}$}
 \label{plot KR wave function1}
\end{center}
\end{figure}

\begin{figure}[!h]
\begin{center}
 \centering
 \includegraphics[width=3.5in,height=2.0in]{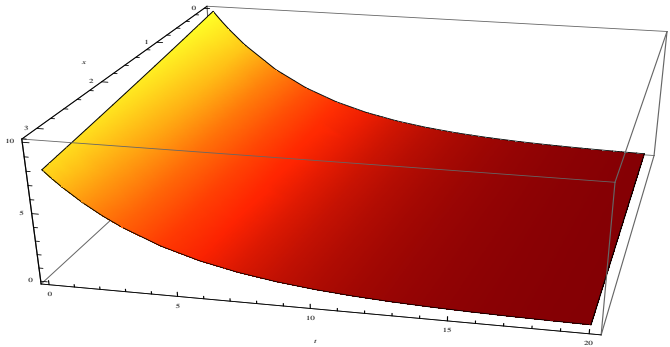}
 \caption{$\chi^{(0)}$ vs $\phi$ (along x axis) and $t$}
 \label{plot KR wave function2}
\end{center}
\end{figure}

Fig.[\ref{plot KR wave function1}] clearly demonstrates that in the regime $\phi \simeq \pi$, the KR wave 
function monotonically decreases with time and 
the decaying time scale ($\tilde{t}\simeq 25$) is less than the exit time of the inflation ($\tilde{t} = 55$). This may explain 
why the present universe carries practically no observable signatures of the rank two antisymmetric Kalb-Ramond field (or 
equivalently the torsion field).\\
Thereby as a whole, the solution of $\chi_v^{(0)}(t,\phi)$ is given by $\chi_v^{(0)}(t,\phi) = e^{-b\phi}f_1(t)$, where $f_1(t)$ is 
obtained in figure[\ref{plot KR wave function1}]. 
Using this solution as a boundary condition, we solve eqn.(\ref{zeroth mode wave function equation}) (evolution of KR wave function 
in the whole bulk) numerically as plotted in Fig.[\ref{plot KR wave function2}].\\

%\begin{figure}[!h]
%\begin{center}
% \centering
% \includegraphics[width=8cm,height=8cm]{bulk_suppression.eps}
 %\caption{$\chi^{(0)}$ vs $\phi$ (along x axis) and $t$}
 %\label{plot KR wave function2}
%\end{center}
%\end{figure}

Fig.[\ref{plot KR wave function2}] reveals that the zeroth mode of KR wave function $\chi^{(0)}(t,\phi)$ decreases with time in 
the whole five dimensional bulk 
i.e for $0 \leq \phi \leq \pi$. However for a fixed $t$, $\chi^{(0)}(t,\phi)$ has different values (in Planckian unit) on hidden and visible brane 
and such hierarchial nature of $\chi^{(0)}(t,\phi)$ (between the two branes) is controlled by the constant $b$.\\
For $T(t) = <T>$, the zeroth mode of KR wave function acquires a constant value throughout the bulk and given by 
\begin{eqnarray}
\chi^{(0)}(t,\phi)\bigg|_{T=<T>} = \sqrt{\frac{k}{<T>}}e^{-k\pi<T>}
\label{constant}
\end{eqnarray}
where we use the normalization condition as shown in eqn.(\ref{wave function normalization}). 
This result is also in agreement with \cite{ssg_prl}. Using the above expression of $\chi^{(0)}(t,\phi)\bigg|_{T=<T>}$, we obtain the coupling 
strengths of Kalb-Ramond field with $U(1)$ gauge field and fermion field on the visible brane as follows \cite{ssg_prl}:
\begin{eqnarray}
 \lambda_{KR-U(1)} = \frac{1}{M_p}e^{-k\pi<T>}
 \label{coupling1}
\end{eqnarray}
and
\begin{eqnarray}
 \lambda_{KR-fer} = \frac{1}{M_p}e^{-k\pi<T>}
 \label{coupling2}
\end{eqnarray}
where $M_p=\sqrt{M^3/k}$. For $k<T> \simeq 12$ (required for solving the gauge hierarchy problem), $e^{-k\pi<T>}$ 
becomes of the order $10^{-16}$. Thereby eqns.(\ref{coupling1}), (\ref{coupling2}) clearly indicate that 
the interaction strengths of KR field to the matter fields are heavily suppressed over the usual 
gravity-matter coupling strength $1/M_p$. This may well serve as an explanation why the large scale behaviour of our 
present universe is solely governed by gravity and carries practically no observable footprints of antisymmetric Kalb-Ramond field.\\

\section{Conclusion}
We consider a five dimensional braneworld model with spacetime torsion caused by a rank-2 
antisymmetric Kalb-Ramond (KR) field in the bulk. The extra spatial dimension 
is $S^1/Z_2$ orbifolded where the orbifolded fixed points are identified with hidden and visible brane respectively. A massive scalar field 
is also considered in the bulk in order to generate a stable potential term for the radion field ($\xi$) - required for stabilizing the interbrane 
separation. We determine the explicit form of the radion potential ($V(\xi)$) as shown in eqn.(\ref{radion potential}). It may 
be observed that $V(\xi)$ goes to zero in absence of the bulk scalar field, as expected \cite{GW}. However the presence of the potential $V(\xi)$ 
activates a dynamics to the radion field governed by the effective field equations. In this scenario, we want to investigate 
whether the dynamics of the radion field can trigger such a dynamical evolution on the KR field, that may lead to an explanation of 
why the effect of torsion is so much weaker than that of curvature on the present visible universe. 
Motivated by these ideas, we solve the cosmological field equations from the perspective of 
four dimensional effective theory. Our findings are as follows:\\
\begin{itemize}
 \item We find that the Kalb-Ramond energy density ($\rho_{KR}$) on our visible universe depends on the on-brane scale factor $a(t)$ as 
 $\rho_{KR} \propto 1/a^6$ (see eqn.(\ref{sol KR energy density})). As we may observe that $\rho_{KR}$ monotonically decreases 
 as the universe expands with time, which leads 
 to a negligible footprint of the KR field on the present universe. However eqn.(\ref{sol KR energy density}) also entails 
 that the energy density of the KR field may be significant in early universe. This points us to explore the dynamical evolution of the 
 KR field from very early phase of the universe. For this purpose, we solve the coupled Freidmann equations 
 for the radion field ($\xi(t)$) and the scale factor ($a(t)$) during initial era and the solutions are given in 
 eqns.(\ref{sol of radion}) and (\ref{sol of scale})) respectively. 
 It is demonstrated in Fig.[\ref{plot brane separation}] that the interbrane separation increases 
 with time and saturates at a constant value ($<T>$) asymptotically. 
 It is also found that without any fine tuning of the parameters, the asymptotic value of the modulus can address the gauge hierarchy problem. 
 On the other hand, the solution of the scale factor corresponds to an accelerating 
 expansion of the early universe and the rate of expansion depends on the parameters $v_v$ and $h_0$ (with $v_v$ and $h_0$ controls 
 the energy density of the bulk scalar field and the KR field respectively). At this stage, it deserves mentioning that 
 in absence of the bulk scalar field ($\Psi$), the radion field becomes constant while the Hubble parameter varies as $H \propto 1/a^3$. 
 This is expected because for $\Psi \rightarrow 0$ (or $v_v = 0$), the potential $V(\xi)$ goes to zero and thus the radion field has 
 no dynamics which in turn makes the variation of the Hubble parameter as $H \propto 1/a^3$ (solely due to the KR field having 
 equation of state parameter $=1$). The duration of inflation ($t_f-t_0$) is obtained in eqn.(\ref{duration}) which 
 reveals that the accelerating phase of the universe terminates within a finite time. Further we also discuss the possible effects of 
 the KR field on the reheating in the present context. We explained that the presence of Kalb-Ramond field makes the reheating time (the time interval after 
 which the production of new particles becomes effective) lesser in comparison to the case when the KR field is absent. However, this is expected 
 because the KR field corresponds to a deceleration of the universe i.e due to the appearance of KR field the Hubble parameter ($H(t)$) 
 decreases with a faster rate by which $H(t)$ reaches to $\Gamma$ (the decay amplitude) more quickly relative to the situation 
 where the KR field is absent.
 
 \item In order to test the model with the observations of $Planck$ 2018 (combining with BICEP-2 Keck-Array data), it is crucial to calculate 
 the spectral index of curvature perturbation ($n_s$) and tensor to scalar ratio ($r$), which are defined in terms of the slow-roll 
 parameter ($\epsilon$). Using these definitions, the expressions of $n_s$ and $r$ are explicitly determined in the present context and 
 as a result, we find that for suitable values of the parameters ($v_v$, $h_0$, $\xi_0$), $n_s$ and $r$ 
 remain within the constraints provided by $Planck$ 2018 \cite{Planck} (see Table[\ref{Table-1}]). Moreover the duration of inflation comes as 
 $10^{-10}$ (GeV)$^{-1}$ if the ratio $m/k$ (bulk scalar field mass to bulk curvature ratio) is taken as $0.2$ \cite{GW}.
 
 \item However the overlap of the zeroth mode KR wave function ($\chi^{(0)}(t,\phi)$) with the visible brane actually 
 fixes the coupling strengths of KR field with various Standard Model fields on the brane. Keeping this in mind, we 
 solve $\chi^{(0)}(t,\phi)$ on the visible brane, numerically, as plotted in Fig.[\ref{plot KR wave function1}]. 
 It is clearly demonstrated 
 that at $\phi=\pi$ the KR wave function monotonically decreases with time and the decaying time scale is less than the exit time of the inflation. 
 Further we also determine the numerical solution for the KR wave function in the whole bulk (see Fig.[\ref{plot KR wave function2}]), 
 which reveals 
 that the effect of $\chi^{(0)}(t,\phi)$ decreases with time in the full five dimensional bulk i.e. for $0 \leq \phi \leq \pi$. 
 However it may be mentioned that the dynamics of $\chi^{(0)}(t,\phi)$ is controlled by the evolution of the radion field 
 and it turns out that for $T(t)=<T>$, $\chi^{(0)}(t,\phi)$ acquires a constant value throughout the bulk as obtained in eqn.(\ref{constant}). 
 Consequently we determine the coupling strengths of KR field with various matter fields on our present visible universe. As a result, 
 such interaction strengths come with a heavily suppressed factor over the usual gravity-matter coupling $1/M_p$. 
 This may provide a natural explanation why the large scale behaviour of our 
present universe is solely governed by gravity and carries practically no observable footprints of spacetime torsion.

\item The second rank antisymmetric Kalb-Ramond field is related to a pseudo-scalar field, known as axion field ($Z(x)$) given by 
$H^{\mu\nu\alpha} = \epsilon^{\mu\nu\alpha\beta}\partial_{\beta}Z$. It may be mentioned that there exist some dark matter models where the axion field was 
considered as a possible candidate to solve the mystery of dark matter \cite{axion1,axion2,axion3}. 
However an experimental program named ABRACADABRA is designed 
to search for axion dark matter and the first results of ABRACADABRA is recently published in \cite{ABRACADABRA} where the authors, through 
estimating the axion-photon coupling, have found no evidence for axion-like cosmic dark matter with 95 percentage C.L. This is consistent with 
the results of our present paper i.e the present universe carries no evidence of axion-like dark matter coming from Kalb-Ramond field.

\end{itemize}

\section*{Appendix-1}
Using eqn.(\ref{em tensor1}) along with the FRW metric, we obtain various components of $T_{\mu\nu}[\xi]$ as follows:
\begin{eqnarray}
 T_{00}[\xi]&=&\frac{1}{2}\dot{\xi}^2 + V(\xi)\nonumber\\
 T_{11}[\xi]&=&T_{22}[\xi] = T_{33}[\xi] = a(t)^2\bigg[\frac{1}{2}\dot{\xi}^2 - V(\xi)\bigg]
 \nonumber
\end{eqnarray}
Further eqn.(\ref{em tensor2}) leads to the various components of $T_{\mu\nu}[B^{(0)}]$ as,
\begin{eqnarray}
 T_{00}[B^{(0)}]&=&\frac{1}{2}\bigg[-h_1h^1 - h_2h^2 - h_3h^3 + h_4h^4\bigg]\nonumber\\
 T_{11}[B^{(0)}]&=&\frac{1}{2}a(t)^2\bigg[h_1h^1 + h_2h^2 - h_3h^3 + h_4h^4\bigg]\nonumber\\
 T_{22}[B^{(0)}]&=&\frac{1}{2}a(t)^2\bigg[h_1h^1 - h_2h^2 + h_3h^3 + h_4h^4\bigg]\nonumber\\
 T_{33}[B^{(0)}]&=&\frac{1}{2}a(t)^2\bigg[-h_1h^1 + h_2h^2 + h_3h^3 + h_4h^4\bigg]
 \nonumber
\end{eqnarray}
\begin{eqnarray}
 T_{10}[B^{(0)}]&=&-h_4h^3~~~~~~~~~~~~,~~~~~~~~~~~T_{20}[H] = h_4h^2\nonumber\\
 T_{30}[B^{(0)}]&=&-h_4h^1~~~~~~~~~~~~,~~~~~~~~~~~T_{12}[H] = a(t)^2h_2h^3\nonumber\\
 T_{13}[B^{(0)}]&=&-a(t)^2h_1h^3~~~~~~,~~~~~~~~~~~T_{23}[H] = a(t)^2h_1h^2
 \nonumber
\end{eqnarray}

\section*{Appendix-2}
The field equation for the zeroth mode Kalb-Ramond field is given by,
\begin{eqnarray}
 \partial_{\mu}\bigg[\sqrt{-g}H^{(0)\mu\nu\lambda}\bigg] = 0
 \label{app2 1}
\end{eqnarray}
where $g$ is the determinant of the on-brane metric. Using the FRW metric ansatz, one obtains $\sqrt{-g} = a^3(t)$, where $a(t)$ 
is the scale factor of the universe. Thus eqn.(\ref{app2 1}) takes the following form,
 \begin{eqnarray}
 &\partial_{\mu}&\bigg[a^3(t)H^{(0)\mu\nu\lambda}\bigg] = 0\nonumber\\
 \Rightarrow &\partial_{0}&\bigg[a^3(t)H^{(0)0\nu\lambda}\bigg] + \partial_{1}\bigg[a^3(t)H^{(0)1\nu\lambda}\bigg]\nonumber\\
 &\partial_{2}&\bigg[a^3(t)H^{(0)2\nu\lambda}\bigg] + \partial_{3}\bigg[a^3(t)H^{(0)3\nu\lambda}\bigg] = 0
 \label{app2 2}
\end{eqnarray}
where the greek indices $\nu$, $\lambda$ run from $0$ to $3$. Therefore for
\begin{itemize}
 \item $\nu = 2$ and $\lambda = 3$, eqn.(\ref{app2 2}) becomes
 \begin{eqnarray}
  &\partial_{t}&\bigg[a^3(t)H^{(0)023}\bigg] + \partial_{x}\bigg[a^3(t)H^{(0)123}\bigg]\nonumber\\
 &\partial_{y}&\bigg[a^3(t)H^{(0)223}\bigg] + \partial_{z}\bigg[a^3(t)H^{(0)323}\bigg] = 0
 \label{app2 3}
 \end{eqnarray}
Due to the antisymmetric nature of KR field, the last two terms of the above equation identically vanish. Further 
from eqn.(\ref{sol off einstein equation}), $H^{(0)023} = 0$. As a result, only the second term of eqn.(\ref{app2 3}) 
survives and leads to the information that the non-zero 
component of KR field ($H^{(0)123}$) is independent of the coordinate $x$ i.e $\partial_{x}\bigg[H^{(0)123}\bigg] = 0$.

\item $\nu = 1$ and $\lambda = 3$, eqn.(\ref{app2 2}) becomes
\begin{eqnarray}
  &\partial_{t}&\bigg[a^3(t)H^{(0)013}\bigg] + \partial_{x}\bigg[a^3(t)H^{(0)113}\bigg]\nonumber\\
 &\partial_{y}&\bigg[a^3(t)H^{(0)213}\bigg] + \partial_{z}\bigg[a^3(t)H^{(0)313}\bigg] = 0
 \label{app2 4}
 \end{eqnarray}
 Here the third term survives, which ensures that $H^{(0)123}$ is independent of $y$.
 
 \item $\nu = 1$ and $\lambda = 2$, eqn.(\ref{app2 2}) becomes
 \begin{eqnarray}
  &\partial_{t}&\bigg[a^3(t)H^{(0)012}\bigg] + \partial_{x}\bigg[a^3(t)H^{(0)112}\bigg]\nonumber\\
 &\partial_{y}&\bigg[a^3(t)H^{(0)212}\bigg] + \partial_{z}\bigg[a^3(t)H^{(0)312}\bigg] = 0
 \label{app2 5}
 \end{eqnarray}
 where the fourth term sustains and gives $\partial_{z}\bigg[H^{(0)123}\bigg] = 0$.
 
\end{itemize}

Therefore it is clear that the non-zero component of the Kalb-Ramond field i.e $H^{(0)123}$ depends only on the time ($t$) coordinate.

\end{document}